\documentclass[journal]{IEEEtran}
\begin{document}

\title{ Construction of Minimal Tail-Biting Trellises for Codes
over Finite Abelian Groups}
\author{Qinqin~Yang
        and~Zhongping~Qin
\thanks{Manuscript received February 3, 2007; revised \ldots
        This work was supported by the National Science Foundation of China under Grant 10571067.}
\thanks{Q. Yang is with the Department of Computer Science, Wuhan University, Wuhan, 430072, China(e-mail:
yqq\_wuda@163.com).}
\thanks{Z. Qin is with the Department  of Software, Huazhong University of
Science and Technology, Wuhan, 430079, China(e-mail:
zpqin0001@126.com).}}
\markboth{IEEE TRANSACTIONS ON INFORMATION THEOREY,~Vol.~1,
No.~11,~November~2007}{}

\maketitle

\begin{abstract}
A definition of atomic codeword for a group code is presented.
Some properties of atomic codewords of group codes are
investigated. Using these properties, it is shown that every
minimal tail-biting trellis for a group code over a finite abelian
group can be constructed from its characteristic generators, which
extends the work of Koetter and Vardy who treated the case of a
linear code over a field. We also present an efficient algorithm
for constructing the minimal tail-biting trellis of a group code
over a finite abelian group, given a generator matrix.
\end{abstract}


\IEEEpeerreviewmaketitle

\newtheorem{theorem}{Theorem}[section]\newtheorem{corollary}{Corollary}[section]
\newtheorem{lemma}{Lemma}[section]\newtheorem{definition}{Definition}[section]
\newtheorem{remark}{Remark}[section]\newtheorem{example}{Example}[section]
\section{Introduction}

\PARstart{T}{rellis } representations of block codes not only
illuminate code structure, but also provide a general framework
for efficient soft-decision decoding of codes [1]-[5], for
instance by using the Viterbi algorithm [6]. Since the decoding
effort is directly related to the complexity of the trellis, such
as the vertex-class profile, the edge-class profile and the
overall Viterbi decoding complexity [7]-[9], characterizing and
constructing minimal trellises for block codes are important in
trellis theory [10]-[22]. It is well known [16-18] that for any
linear code over a field or group code, there exists a unique
minimal conventional trellis up to isomorphism. Furthermore, the
minimal conventional trellis for any linear code over a field can
be easily constructed from its generator matrix or parity check
matrix by several methods [15] and [19]. In [20], the authors
proved that the minimal trellis for a group code over a finite
abelian group is the product of some minimal trellises for linear
codes over $Z_{p^\alpha}$, then for arbitrary finite abelian
group, it is sufficient to consider a linear code over a ring
$Z_{p^\alpha}$. By the method, they generalized the result in [15]
into codes over finite abelian groups and proved that the minimal
conventional trellis for any group code can be easily constructed
from its biproper $p$-basis. A tail-biting trellis may be has less
complexity than the minimal conventional trellis for the code,
[23] and [24]. However, much less is known about efficient
construction of tail-biting trellises. The authors in [12] proved
that any linear tail-biting trellis for a linear code over a field
can be constructed from its characteristic generators, and
conjectured that it is correct for a group code over a finite
abelian group. In [25], the authors presented the difficulty when
they try to prove the conjecture in [12].

In this paper, our goal is to show that the conjecture in [12] is
true and present an algorithm for constructing the minimal
tail-biting trellis of a group code over a finite abelian group,
given a generator matrix. A key step toward proving the conjecture
is handling codes over cyclic groups $C_{p^\alpha}$. Such a code
can be viewed as a linear code over ring $Z_{p^\alpha}$. Because
the order of $p$-generator sequence, $p$-bases of a linear code
over $Z_{p^\alpha}$ do not share the useful properties of a basis
of a linear code over a field. We get around this difficulty by
introducing the notion of atomic codeword of a linear code over
$Z_{p^\alpha}$, which enjoys properties similar to those of atomic
codeword for a linear code over a field.

We start with the definitions of conventional and tail-biting
trellises in section II. We then introduce a number of concepts
related to tail-biting trellises. It follows from [20] that the
problem of constructing minimal tail-biting trellises of a block
code over a finite abelian group reduces to the case of linear
codes over ring $Z_{p^\alpha}$. Thus, we introduce the concepts of
$p$-generator sequences and $p$-linear combinations of a linear
code over $Z_{p^\alpha}$.

In section III, we present a rigorous definition of atomic
codeword for a group code. Some properties of atomic codewords of
group codes are investigated. It is well known in [15] that
minimal conventional trellises for linear codes over fields are
obtained by forming the product of elementary trellises
corresponding to the one-dimensional subcodes generated by atomic
codewords and the structure of the trellis is determined solely by
the spans of the atomic codewords. These spans, called atomic
spans, are uniquely determined by the linear code over a field.
Furthermore, any set of $k$ codewords of the linear code of
dimension $k$ over a field with atomic spans is a basis in
minimal-span form. Any two bases in minimal-span form give the
same set of conventional spans, although a basis in minimal-span
form is not unique. We can find a biproper $p$-basis of a linear
code over $Z_{p^\alpha}$ shares the useful properties of a basis
in minimal-span form of a linear code over a field. This make us
to define a definition of atomic codeword for a group code and it
is possible to extend the work of Koetter and Vardy.

In section IV, using these properties of atomic codewords of group
codes, we show the conjecture in [12] is true. We also proved that
the conjecture in [12] is true under other minimality orders for
tail-biting trellis. Therefore, we show that although minimal
tail-biting trellises for group codes are generally not unique,
every minimal linear tail-biting trellis for a group code over a
finite abelian group necessarily can be construct from its
characteristic matrix. This gives a general solution to the
problem of constructing minimal linear tail-biting trellises for
group codes over finite abelian groups.

In section V, we consider codes over finite abelian groups. Since
a code over a finite abelian group can decomposed into a direct
product of codes over those abelian $p$-groups which are all Sylow
$p$-subgroup of the group and a code of length $n$ over $p$-group
is equivalent to a linear code
   of length $mn$ over $Z_{p^{\alpha}}$, for arbitrary finite abelian
group, it is sufficient to consider the case of a linear code over
a ring $Z_{p^\alpha}$. Therefore, we present an algorithm for
computing this characteristic matrix in time $\emph{O}(n^3)$ from
any $p$-basis for a linear code over a ring $Z_{p^\alpha}$. For
arbitrary finite abelian group, using sectionalization, we can
obtain a minimal linear tail-biting trellis for a code over the
group from minimal linear tail-biting trellises for linear codes
over $Z_{p^\alpha}$, which is similar to the method in [20]. Thus
we get an efficient algorithm for constructing the minimal
tail-biting trellis of a group code over a finite abelian group,
given a generator matrix.

The research on trellises for lattices also an important topic.
The problem of constructing minimal trellises is still open. Since
this problem essentially reduces to that of constructing minimal
trellises for block codes over abelian groups, [26] and [27], the
most important application of our work is to the construction of
minimal trellises for lattices. In [28], the authors proved that
under the Gray map from $(Z_2)^2$ to the ring $Z_4$, these codes
are linear over $Z_4$, or equivalently, are group codes over
$C_4$. Therefore, another application is to the construction of
minimal trellises for some famous nonlinear binary codes,
including Kerdock, Preparata, and Goethals codes.

\section{Preliminaries}

In this section, We first introduce some definitions that will be
used in the paper, see also [29].

 Let $G$ be a finite abelian group. A subgroup $C$ of
$G^n$ under the componentwise addition operation of $G$ is said to
be \emph{a group code over $G$}. Let $R$ be a ring . $C$ is\emph{
a linear code over $R$} if $C$ is a subgroup of $R^n$ under the
componentwise addition operation of $R$ and $C$ is closed under
componentwise multiplication with elements of $R$. In fact, a
linear code $C$ over $R$ is exactly a submodule of $R^n$. When
$R=F_q$, a finite field, then $C$ is a linear code over $F_q$.
Clearly, the class of linear codes over fields is contained in the
class of linear codes over rings, which is in turn contained in
the class of group codes.

  Now we introduce some basic concepts
on conventional and tail-biting trellises.

 An edge-labeled
directed graph is a triple $(V,E,A)$, consisting of a set $V$ of
vertices, a finite set $A$ called the alphabet, and a set $E$ of
ordered triples $(v,a,v')$, with $v,v'\in V$ and $a\in A$ called
edges. We say that an edge $(v,a,v')\in E$ begins at $v$, ends at
$v'$, and has label $a$.\\

 \begin{definition}
A \emph{conventional trellis} $T=(V,E,A)$ of length $n$ is an
edge-labeled directed graph with the following property: the
vertex set $V$ can be partitioned as
\begin{equation}
 V=V_0\cup V_1\cup\cdots \cup V_n
 \end{equation}
  where $|V_0|=|V_n|=1$, such that every edge in $T$ begins at
a vertex of $V_i$ and ends at a vertex of $V_{i+1}$, for some
$i=0,1,\ldots, n-1$. The ordered index set
${\mathcal{I}}=\{0,1,\ldots,n\}$ induced by the partition in (1)
is called the \emph{time axis} for $T$.\\
\end{definition}

The trellis $T$ is \emph{reduced} if every vertex in $T$ lies on
at least one path from a vertex in $V_0$ to a vertex in $V_n$. The
trellis $T$ said to \emph{represent} a block code $C$ of length
$n$ over $A$ if $C$ is precisely the set of all edge-labeled
sequences corresponding to those paths in $T$ that start at a
vertex of $V_0$.

In the following, we will see that tail-biting trellises may be
viewed as a generalization of a conventional trellis to a circular
time axis.\\

\begin{definition}
A \emph{tail-biting trellis} $T=(V,E,A)$ of length \\
$n$ is an edge-labeled directed graph with the following property:
the vertex set $V$ can be partitioned as
\begin{equation}
V=V_0\cup V_1\cup\cdots \cup V_{n-1}
 \end{equation}
such that every edge in $T$ begins at a vertex of $V_i$ and ends
at a vertex of $V_{i+1}$, for some $i=0,1,\ldots, n-2$, or begins
at a vertex of $V_{n-1}$ and ends at a vertex of $V_0$. The
ordered index set ${\mathcal{I}}=\{0,1,\ldots,n-1\}$ induced by
the partition in (2)
is called the \emph{time axis} for $T$.\\
\end{definition}

An index interval $[i,j]$ represents the sequence
$\{i,i+1,\ldots,j\}$ if $i\leq j$, and the sequence
$\{i,i+1,\ldots,n-1,0,\ldots,j\}$ if $i>j$. Such interval $[i,j]$
is called closed \emph{cycle interval}, we let semiopen cyclic
interval $(i,j]$ denote $[i,j]\setminus \{i\}$. The tail-biting
trellis $T$ is \emph{reduced} if every vertex in $T$ lies on at
least one cycle from a vertex in $V_0$ to a vertex in $V_n$.  The
tail-biting trellis $T$ said to \emph{represent} a block code $C$
of length $n$ over $A$ if $C$ is precisely the set of all
edge-labeled sequences corresponding to those
 cycles in $T$ that start at a vertex of $V_0$.

  Let $C(T)$ denote the code represented by a trellis $T$,
either conventional or tail-biting. Let $T=(V,E,A)$ be a trellis,
either conventional or tail-biting, of length $n$. The ordered
sequence
$$\Theta(T)=(|V_0|,|V_1|,\ldots,|V_{n-1}|)$$
 is called
the \emph{vertex-class profile} of $T$. For a given code $C$, we
say that a trellis $T$ is less than or equal to another trellis
$T'$, denoted as $T\preceq_{\Theta}T'$, if
\begin{equation}
|V_i|\leq|V'_i|, \;\mbox{for all} \; i=0,1,\ldots,n-1
\end{equation} If there is at least a
strict inequality, then $T$ is strictly less than $T'$ and write
$T\prec_{\Theta}T'$.\\

\begin{definition}
A trellis $T$, either conventional or tail-biting, for a code $C$
of length $n$ is \emph{minimal under $\prec_{\Theta}$}, or simply
\emph{minimal}, if there does not exist a trellis $T'$ such that
$T'\prec_{\Theta}T$.\\
\end{definition}

It follows from [16], [17] and [20] that for linear or group codes
there exists a unique minimal conventional trellis under
$\prec_{\Theta}$ up to graph isomorphism. The unique minimal
trellis has the smallest possible vertex count simultaneously at
all times and minimizes all conceivable measures of trellis
complexity [18]. But there are many incomparable minimal
tail-biting trellises under $\prec_{\Theta}$ up to graph isomorphism.

Now we introduce several total orders under which any two
trellises are comparable. In the following, we let $E_i$ denote
the set of edges that start in a vertex of $V_i$ and end in a
vertex of $V_{i+1}$.

\begin{eqnarray}
\mbox{\bf product order:}\;\;\;\;\; T\preceq_{\Pi}
T'\;\mbox{if}\;\prod_{i=0}^{n-1}|V_i|\leq
\prod_{i=0}^{n-1}|V'_i| \\
 \mbox{\bf max order:}\;\;\;\;\; T\preceq_{max}
T'\;\mbox{if}\;\max\limits_{i}|V_i|\leq \max\limits_i|V'_i|\\
 \mbox{\bf vertex-sum order:}\;\;\;\;\; T\preceq_{\Sigma}
T'\;\mbox{if}\;\sum_{i=0}^{n-1}|V_i|\leq \sum_{i=0}^{n-1}|V'_i|\\
 \mbox{\bf edge-product order:}\;\;\;\;\; T\preceq_{ \Pi\mathcal{E}}
T'\;\mbox{if}\;\prod_{i=0}^{n-1}|E_i|\leq
\prod_{i=0}^{n-1}|E'_i|\\
\mbox{\bf edge-max order:}\;\;\;\;\; T\preceq_{max\mathcal{E}}
T'\;\mbox{if}\;\max\limits_{i}|E_i|\leq \max\limits_i|E'_i|\\
 \mbox{\bf edge-sum order:}\;\;\;\;\; T\preceq_{\Sigma\mathcal{E}}
T'\;\mbox{if}\;\sum_{i=0}^{n-1}|E_i|\leq \sum_{i=0}^{n-1}|E'_i|
\end{eqnarray}

 It is obvious that if $T\preceq_{\Theta}T'$ implies that $T\preceq_{\Pi}
 T'$, $T\preceq_{max}T'$ and $T\preceq_{\Sigma} T'$. Then the set
 of $\preceq_{\Pi}$-minimal trellises, the set of $\preceq_{max}$-minimal trellises and
 the set of $\preceq_{\Sigma}$-minimal trellises for a given code
 are subsets of  the set of $\preceq_{\Theta}$-minimal trellises for the same
 code. On the other hand, the set of $\preceq_{\Pi\mathcal{E}}$-minimal linear trellises for a given code
 are subsets of  the set of $\preceq_{\Theta}$-minimal linear trellises for the same
 code. But this is not true for the edge-max order or the edge-sum
 order, see [12]. \\

\begin{definition}
Let $T'=(V',E',A)$ and $T''=(V'',E'',A)$ be two trellises of
length $n$, either conventional or tail-biting, over the alphabet
$A$, and assume that $A$ is endowed with an associative addition
operation. Then the \emph{product trellis } $T'\times T''$ is the
trellis $T=(V,E,A)$ whose vertex classes and edge classes are the
Cartesian products, defined as follows:
\begin{eqnarray}
V_i=\{(v',v''):v'\in V'_i \;\mbox{and} \; v''\in V''_i\}\ & & \\
E_i=\{((v'_1,v''_1),a'+a'',(v'_2,v''_2)):\quad \ \ & & \nonumber\\
\quad\qquad(v'_1,a',v'_2)\in E'_i
\;\mbox{and}\;(v''_1,a'',v''_2)&\in E''_i\}.\qquad\quad&
\end{eqnarray}
 There is an edge $e\in
E_i$ in $T$ labeled $a$, from a vertex $(v'_1,v''_1)\in V_{i-1}$
to a vertex $(v'_2,v''_2)\in V_i$ if and only if
$(v'_1,a',v'_2)\in E'_i$, $(v''_1,a'',v''_2)\in E''_i$ and
$a=a'+a''$. If $C_1=C(T')$ and $C_2=C(T'')$, then the product
trellis $T=T\times T'$ represents the code
\begin{eqnarray}
C_1+C_2=\{c_1+c_2: c_1\in C_1, c_2\in C_2\}.
\end{eqnarray}
Clearly, the trellis product operator
is both associative and commutative.\\
\end{definition}

 Notice that in the paper, a trellis is either conventional
or tail-biting. We will see that the notion of a trellis product
and the corresponding product construction are very important for
constructing minimal trellises, see also [12-16] and [18]-[21].

If $C$ is a linear code of length $n$ and dimension $k$ over the
field $F_q$, let $\{x_1,x_2,\ldots,x_k\}$ be a basis for $C$ and
let $\langle x_i\rangle$ denote the one-dimensional subcode of $C$
generated by $x_i$. Then $C=\langle x_1\rangle+\langle
x_2\rangle+\cdots+\langle x_k\rangle$. If $T_1,T_2,\ldots,T_k$ are
trellises for $\langle x_1\rangle,\langle
x_2\rangle,\ldots,\langle x_k\rangle$, respectively, then their
product $T=T_1\times T_2\times \cdots\times T_k$ represents $C$.
This completes the description of the product construction if we
can specify the trellises $T_1,T_2,\ldots,T_k$.

 Now we introduce the notions of span and
elementary trellis. Given a codeword $x\in C$, a \emph{span} of
$x$, denote $[x]$, is a semiopen interval $(i,j]\in\mathcal{I}$
such that the corresponding closed interval $[i,j]$ contains all
the nonzero positions of $x$. We also allow $[x]=\mathcal{I}$. For
a codeword $x$ with span $(i,j]$, an \emph{elementary trellis} of
$x$, denote $T_x$, is the minimal trellis for $\langle x\rangle$
which is a one-dimensional code generated by $x$.

 Given a codeword $x=(x_0,x_1,\ldots,x_{n-1})$ over the field
$F_q$, along with its span $[x]=(a,b]$, the corresponding
elementary trellis $T_x$ can be constructed as follows: $T_x$ has
$q$ vertices labeled by the elements of $F_q$ at times
$a+1,\ldots,b$, and a single vertex $0$, at other positions.
 There is an edge $e\in E_i$ from a vertex $v\in V_i$ to a vertex $v'\in V_{i+1}$
 if and only if $i=a$, or $i=b$, or the two vertices $v,v'$ have the same label.
 All the edge-label sequences in $T_x$ are the $q$ different multiples of $x$. Note
 that an elementary trellis $T_x$ depends not only on $x$ and span $(a,b]$, but also
 on the ambient field $F_q$.

 Similarly, for a codeword $x$ of a group code, along with its
span $[x]=(a,b]$ and its order $q$, the corresponding elementary
trellis $T_x$ can be constructed as follows: $T_x$ has $q$
vertices labeled by the elements of group $\langle x\rangle$ at
times $a+1,\ldots,b$, and a single vertex $0$, at other positions.
 There is an edge $e\in E_i$ from a vertex $v\in V_i$ to a vertex $v'\in V_{i+1}$
 if and only if $i=a$, or $i=b$, or the two vertices $v,v'$ have the same label.
 All the edge-label sequences in $T_x$ are the $q$ different multiples of
 $x$.

  From now on, we search for a solution to the problem of
constructing minimal tail-biting trellises of a block code over a
finite abelian group. It follows from [20] that the minimal
trellis for group code
    $C$ over a finite abelian group is the product of some minimal trellises for linear codes
over $Z_{p^\alpha}$. Thus we focus on our discussion on linear
codes over $Z_{p^\alpha}$.

 Now we will introduce a number of concepts related to linear
codes over $Z_{p^\alpha}$. Let $V=\{x_1,x_2,\ldots,x_k\}$ be a set
of codewords of a linear code $C$ over $Z_{p^\alpha}$.
$\sum\limits_{i=1}^k a_ix_i$ is a \emph{$p$-linear combination }of
these codewords if all coefficients $a_i\in Z_{p}$. Denoted by
\emph{$p$-span($V$) }the set of all elements generated by
$p$-linear combination of the elements in $V$. An ordered sequence
of $\{x_1,x_2,\ldots,x_k\}$ over $Z_{p^\alpha}$ is said to be a
\emph{$p$-generator sequence} if for $1\leq i\leq k$, $px_i$ is a
$p$-linear combination of the codewords $x_{i+1},\ldots,x_k$ (in
particular, $px_k=0$). For any $p$-generator sequence $V$, it was
proved that $p$-span($V$)=span($V$) and that the codeword $0$ is a
nontrivial $p$-linear combination of these codewords in $V$ if and
only if there is a codeword in $V$ that can be expressed as a
$p$-linear combination of the remaining codewords in $V$. A
$p$-generator sequence $V$ is \emph{$p$-linearly independent} if
the codeword $0$ can not be expressed as a nontrivial $p$-linear
combination of these codewords in $V$. A $p$-linearly independent
$p$-generator sequence $V$ is called a \emph{$p$-basis of
$p$-span($V$)}. Obviously, the $p$-linear combination of the
codewords of $p$-basis $V$ uniquely generate the elements of the
module $p$-span($V$). If $|V|=k$, then the module has $p^k$
elements and we say that the \emph{$p$-dimension} of the module is
$k$. We say that $u$ and $v$ in $Z_{p^\alpha}$ are associates if
there exists a unit $w$ in $Z_{p^\alpha}$ such that $u=wv$. Then
$u$ and $v$ are associates if and only if they have the same
order. A $p$-generator sequence $V$ is \emph{proper} if for any
two codewords $u,v\in V$, either $u$ and $v$ start at different
positions, or they start the same position but their starting
components are not associates. Similarly, a $p$-generator sequence
$V$ is \emph{coproper} if for any two codewords $u,v\in V$, either
$u$ and $v$ end at different positions, or they end at the same
position but their ending components are not associates. A
$p$-generator sequence $V$ is \emph{biproper} if it is proper and
coproper. A proper $p$-generator sequence
$V=\{x_1,x_2,\ldots,x_k\}$ is in \emph{row echelon form} if for
$1\leq i<j\leq k$, either $v_i$ has an earlier starting position
than $v_j$, or $v_i$ and $v_j$ have the same starting position but
the starting component of $v_i$ has higher order than the starting
component of $v_j$. By Theorem 6.12 and Lemma 7.1 in [20], we can
get a proper $p$-generator sequence in row echelon form from any
$p$-generator sequence and a biproper $p$-basis in row echelon
form from any proper $p$-generator sequence in row echelon form.
Any submodule of $Z_{p^\alpha}^n$ has a biproper $p$-basis by
Theorem 6.11 in [20].

 Therefore, for a codeword $x$ over $Z_{p^\alpha}$, along with
its span $[x]=(a,b]$, an elementary trellis of $x$ over
$Z_{p^\alpha}$, denote $T_x$, is the minimal trellis for
$p$-span($\{x\}$) and can be constructed as follows: $T_x$ has $p$
vertices labeled by the elements of $Z_p$ at times $a+1,\ldots,b$,
and a single vertex $0$, at other positions.
 There is an edge $e\in E_i$ from a vertex $v\in V_i$ to a vertex $v'\in V_{i+1}$
 if and only if $i=a$, or $i=b$, or the two vertices $v,v'$ have the same label.
 All the edge-label sequences in $T_x$ are the $p$ different multiples of $x$.
Let the order of $x$ is $p^{r}$. The minimal trellis $T$ for
$\langle x\rangle$ which is a cyclic code generated by $x$ over
$Z_{p^\alpha}$ and can be constructed as $T=T_{x}\times
T_{px}\times\cdots,\times T_{p^{r-1}x}$.

\section{  Some results on atomic classes}
In this section, we present a definition of atomic codeword for a
linear code over $Z_{p^\alpha}$, which is a generalization of the
notion of atomic codeword of a linear code over $F_q$, and study
their basic structural properties.

 For a codeword
$x=(x_0,x_1,\ldots,x_n)\in C$, we let $\triangleleft(x)$ denote
the smallest integer $i$ such that $x_i\neq0$, $\triangleright(x)$
denote the largest integer $j$ such that $x_j\neq0$,  and $
o_1(x)$ and $o_2(x)$ denote the orders of the first and last
nonzero components of $x$, respectively. We say that
$(\triangleleft(x),\triangleright(x)]$ is the\emph{ conventional
span }of $x$ and $((\triangleleft(x),\triangleright(x)], o_1(x),
o_2(x))$ is \emph{a characteristic triple of $x$}. The \emph{span
length} of $x$ is defined as
$\triangleright(x)-\triangleleft(x)+1$. In the following, We will
use the concept of characteristic triple of codeword to define an
equivalence relation on any linear code $C$
 over $Z_{p^\alpha}$.\\

\begin{definition}
Two codewords of a linear code $C$
 over $Z_{p^\alpha}$ are \emph{equivalent} if and only if they
 have the same characteristic triple.
An \emph{atomic class} of a linear code $C$
 over $Z_{p^\alpha}$ is an equivalence class that
their elements cannot be expressed as $p$-linear combinations of
codewords from $C$ of strictly smaller span lengths or $p$-linear
combinations of a codeword having strictly smaller span length and
a codeword having the same conventional span and strictly smaller
order of the first or the last nonzero component. The elements of
an atomic class are called \emph{atomic codewords}.
\end{definition}

\medskip
\begin{remark}
Obviously, any multiple of an atomic codeword by a unit in
$Z_{p^\alpha}$ is also an atomic codeword with the same
characteristic triple; however, an atomic class may consist of
multiples of more than one codeword by units in $Z_{p^\alpha}$.
Note that the elements of a biproper $p$-basis of a linear code
over $Z_{p^\alpha}$ are also atomic codewords.

For a field $F_q$, the order of any nonzero element in
the addition group of $F_q$ is the character of $F_q$. Thus, the above definition is
 a generalization of the definition of atomic codeword of a linear code over $F_q$.
\end{remark}

\medskip
In the following, we investigate some properties of atomic
codewords.\\

 \begin{theorem}
 If $c_1$ and $c_2$ are atomic codewords in linear code $C$
 over $Z_{p^\alpha}$ and $(\triangleleft(c_1), o_1(c_1))\neq(\triangleleft(c_2),
 o_1(c_2))$ or $(\triangleright(c_1), o_2(c_1))\neq(\triangleright(c_2),
 o_2(c_2))$, then $(\triangleleft(c_1), o_1(c_1))\neq(\triangleleft(c_2),
 o_1(c_2))$ and $(\triangleright(c_1), o_2(c_1))\neq(\triangleright(c_2),
 o_2(c_2))$.
 \end{theorem}
\begin{proof}
Suppose that for $c_1$ and $c_2$, $(\triangleleft(c_1),
o_1(c_1))=(\triangleleft(c_2),
 o_1(c_2))$ and $(\triangleright(c_1), o_2(c_1))\neq(\triangleright(c_2),
 o_2(c_2))$. Then $\triangleright(c_1)\neq\triangleright(c_2)$ or
  $\triangleright(c_1)=\triangleright(c_2)$ and $o_2(c_1)\neq
  o_2(c_2)$. Without loss of generality, let
  $\triangleright(c_1)<\triangleright(c_2)$ or $\triangleright(c_1)=\triangleright(c_2)$ and $o_2(c_1)<
  o_2(c_2)$. Then there exists a unit $\alpha$ in $Z_{p^\alpha}$ such
  that $c_3=c_2-\alpha c_1$ starts later than $c_2$. Clearly,
   $\alpha c_1$ is also an atomic codeword with the same characteristic triple as $c_1$. Now
  $c_2=c_3+\alpha c_1$, therefore $c_2$ can be expressed as
  a $p$-linear combination of a codeword $c_3$ of strictly smaller span length
  and a codeword  $\alpha c_1$ having either strictly smaller span length or the same conventional span and strictly smaller
order of the last nonzero component, contradicting the assumption
that $c_2$ is atomic. The case in which the codewords have
$(\triangleleft(c_1), o_1(c_1))\neq(\triangleleft(c_2),
 o_1(c_2))$ and $(\triangleright(c_1), o_2(c_1))=(\triangleright(c_2),
 o_2(c_2))$ is proved similarly.
 \end{proof}

\medskip
\begin{theorem}
 Let $C$ be a linear code of length
$n$ and $p$-dimension $k$ over $Z_{p^\alpha}$. Then the elements
of any set $A$ of atomic codewords with the property that no two
members of $A$ belong to the same atomic class are $p$-linearly
independent.
 \end{theorem}
\begin{proof}
Any set of codewords, no two of which have the same starting
position and order of the first nonzero component, are
$p$-linearly independent and the elements of $A$ have this
property.
\end{proof}

\medskip
\begin{theorem}
Let $C$ be a linear code of length $n$ and $p$-dimension $k$ over
$Z_{p^\alpha}$. Then every codeword $c$ in $C$ can be expressed as
a $p$-linear combination of atomic codewords, each from a
different atomic class. Moreover, a complete set of atomic class
representatives in row echelon form is a biproper $p$-basis for
$C$ in row echelon form. Therefore, code $C$ has $k$ distinct
atomic classes.
\end{theorem}
\begin{proof}
It is trivial if $c$ is atomic. Now suppose $c$ is not atomic,
then $c$ can be be expressed as a $p$-linear combination of
codewords from $C$ of strictly smaller span lengths or a
$p$-linear combination of a codeword having strictly smaller span
length and a codeword having the same conventional span and
strictly smaller order of the first or the last nonzero component.
Any combination of two codewords having the same characteristic
triple may be replaced by either a single codeword with the same
characteristic triple, or a codeword of strictly smaller span
length. If any terms in this combination are themselves not
atomic, then they can be further expressed a $p$-linear
combination of codewords of strictly smaller span lengths or a
$p$-linear combination of a codeword having strictly smaller span
length and a codeword having the same conventional span and
strictly smaller order of the first or the last nonzero component.
Continuing in this way, we obtain a chain of strictly decreasing
nonnegative span lengths, which is bound to terminate in finite
steps, with the result that $c$ is expressed as a $p$-linear
combination of atomic codewords, no two from the same atomic
class. Let $\{x_1,x_2,\ldots,x_m\}$ be a complete set of atomic
class representatives in row echelon form. We will show that
$\{x_1,x_2,\ldots,x_m\}$ is a $p$-generator sequence. For $1\leq
i\leq m$, $px_i$ is a codeword starting later than $x_i$ or having
the same starting position as $x_i$ and strictly smaller order of
the first nonzero component than $x_i$, then $px_i$ is expressed
as a $p$-linear combination of atomic codewords
$x_{i+1},\ldots,x_m$ (in particular, $px_m=0$). By Theorem 3.1 and
Theorem 3.2, $\{x_1,x_2,\ldots,x_m\}$ is a biproper $p$-basis for
$C$ in row echelon form. Therefore, $m=k$ and code $C$ has $k$
distinct atomic classes.
\end{proof}

\medskip
In general, for a linear code $C$ over $Z_{p^\alpha}$, a biproper
$p$-basis in row echelon form is not unique. However, by Theorem
3.3 and Remark 3.1, we have the following.\\

 \begin{corollary}
Any two biproper $p$-bases in row echelon form have the same set
of conventional spans and orders of the first and last nonzero
components of the elements of biproper $p$-basis, which are
uniquely determined by $C$.
\end{corollary}

\medskip
\begin{theorem}
Let $C$ be a linear code of length $n$ and $p$-dimension $k$ over
$Z_{p^\alpha}$. Then any codeword $c$ with the same characteristic
triple as an atomic codeword $a$ is atomic.
\end{theorem}
\begin{proof}
If $c$ were not atomic, $c$ could be expressed as a $p$-linear
combination of atomic codewords strictly smaller span lengths or a
$p$-linear combination of atomic codewords of strictly smaller
span lengths and a atomic codeword having the same conventional
span and strictly smaller order of the first or the last nonzero
component, one of which, $b$ say, would start in the same position
and have the same order of the first nonzero component as $c$.
Therefore $b$ have the same starting position and order of the
first nonzero component as $a$, which contradict the above Theorem
3.1.
\end{proof}

\medskip
Therefore, we have the following.\\

 \begin{corollary}
 Any set of $k$ codewords
of a linear code $C$ of length $n$ and $p$-dimension $k$ over
$Z_{p^\alpha}$ in row echelon form with the same set of
characteristic triples as a biproper $p$-basis of $C$ is also a
biproper $p$-basis for $C$ in row echelon form.
 \end{corollary}

\medskip
\begin{theorem}
Let $C$ be a linear code of length $n$ and $p$-dimension $k$ over
$Z_{p^\alpha}$. Then the $p$-dimension of any subcode $S$ of $C$
in which every codeword $c$ has a conventional span contained in
$(a,b]$ and the maximal orders of the $a$-th and the $b$-th
components of codewords are respectively $s$ and $t$, is equal to
the number of atomic classes in $C$ whose conventional span is in
$(a,b]$ and order of the $a$-th component is not higher than $s$
and order of the $b$-th component is not higher than $t$.
\end{theorem}
\begin{proof}
 Every codeword $c$ in $S$ can be
expressed as a $p$-linear combination of atomic codewords from
different atomic classes. Then in the $p$-linear combination, no
atomic codeword with conventional span not in $(a,b]$ or whose
conventional span is in $(a,b]$ and order of the $a$-th component
is higher than $s$ or order of the $b$-th component is higher than
$t$ can be used. Therefore, we finish the proof.
 \end{proof}

\medskip
\begin{theorem}
 Let $C$ be a linear code of length
$n$ and $p$-dimension $k$ over $Z_{p^\alpha}$. Then the set of
codeword characteristic triples achieved by $C$ is completely
determined by the set of characteristic triples of atomic classes
in the code.
\end{theorem}
\begin{proof}
Every codeword $c$ in $C$ can be expressed as a $p$-linear
combination $\sum_j a_jx_j$ where $a_j\in Z_p$, $a_j\neq0$ and
$x_j$ are atomic with characteristic triple
$((\triangleleft(x_j),\triangleright(x_j)],o_1(x_j),\\o_2(x_j))$.
Then the characteristic triple of $c$ is
\begin{eqnarray*}
((\min\limits_j\{\triangleleft(x_j)\},\max\limits_j\{\triangleright(x_j)\}],&
\max\limits_{\triangleleft(x_m)=\atop\min\limits_j\{\triangleleft(x_j)\}}\{o_1(x_m)\},&\\
\max\limits_{\triangleright(x_t)=\atop\max\limits_j\{\triangleright(x_j)\}}\{o_2(x_t)\}),\qquad&
 &
\end{eqnarray*}
and hence is determined by the set of  characteristic triples of
atomic classes.
 \end{proof}

\medskip
\begin{theorem}
Let $C$ be a linear code of length $n$ and $p$-dimension $k$ over
$Z_{p^\alpha}$. Then any equivalence class of minimum nonzero span
length is atomic and consists of multiples of a single codeword by
units in $Z_{p^\alpha}$.
\end{theorem}
\begin{proof}
If an element of the equivalence class of minimum nonzero span
length is not atomic, it can be  expressed as a $p$-linear
combination of codewords of strictly smaller span lengths or a
$p$-linear combination of a codeword having strictly smaller span
length and a codeword having the same conventional span and
strictly smaller order of the first or the last nonzero component,
which is a contradiction. Suppose two codewords $c_1$ and $c_2$
belong to the same equivalence class of minimum nonzero span
length. Then all  multiples of $c_1$ and $c_2$ by units in
$Z_{p^\alpha}$
 also belong to the same atomic class. There exists a unit $\alpha$
  in $Z_{p^\alpha}$ such that $c_1-\alpha c_2$ starts later than
  $c_1$. If $c_1-\alpha c_2\neq0$, then $c_1-\alpha c_2$ is a codeword
  of smaller span length than $c_1$.
  Therefore $c_1-\alpha c_2=0$, which implies  $c_1=\alpha c_2$.
 \end{proof}
\section{construction of minimal tail-biting trellises}
In this section, we prove the conjecture in [12] is true that
every minimal tail-biting trellis for a group code over a finite
abelian group can be constructed from its characteristic
generators, which extends the work of Koetter and Vardy who
treated the case of a linear code over a field. Since for
arbitrary finite abelian group, it is sufficient to consider the
case of a linear code over a ring $Z_{p^\alpha}$. Thus we first
focus on our discussion on linear codes over $Z_{p^\alpha}$.\\

 \begin{theorem}
 Let $X=\{x_1,x_2,\ldots,x_k\}$ be a $p$-basis for a linear
code $C$ over $Z_{p^\alpha}$ and $Y=\{y_1,y_2,\ldots,y_k\}$ be a
biproper $p$-basis for $C$ in row echelon form. Let $T=T_{x_1}
   \times\cdots\times T_{x_k}$ is a $\prec_{\Theta}$ minimal
linear tail-biting trellis for $C$. Then for each $1\leq i\leq k$,
either $0\in[x_i]$ or there exists $y\in Y$ such that $x_i$
belongs to the same equivalence class of $C$, namely,  $([x_i],
o_1(x_i),o_2(x_i))=((\triangleleft (y),\triangleright (y)],
o_1(y),o_2(y))$.
\end{theorem}
\begin{proof}
We proceed by induction on $p$-dimension $k$ of $C$. If $k=1$ the
result is clear by Theorem 3.3. Suppose that $k>1$. It is easy to
show that $\{x_2,\ldots,x_k\}$ is a $p$-linearly independent
$p$-generator sequence. Let $C'=p$-span($\{x_2,\ldots,x_k\}$) be a
subcode of $C$. It is obvious that $T'=T_{x_2}
   \times\cdots\times T_{x_k}$ is a $\prec_{\Theta}$ minimal
   linear tail-biting trellis for $C'$. By definition the $p$-dimension of
$C'$ is $k-1$ and $\{x_2,\ldots,x_k\}$ is a $p$-basis for $C'$.
Therefore, every codeword in $C'$ can be expressed as a linear
combination of atomic codewords from $k-1$ distinct atomic classes
of $C$ by Theorem 3.5. Let $Z'=\{z_2,\ldots,z_k\}$ be a complete
set of representatives of the $k-1$ distinct atomic classes in row
echelon form. By Theorem 3.3, $Z'=\{z_2,\ldots,z_k\}$ is a
biproper $p$-basis for $C'$ in row echelon form. Then by induction
for each $2\leq i\leq k$, either $0\in[x_i]$ or there exists $z\in
Z'$ such that $x_i$ belongs to the same equivalence class of $C$,
namely, $([x_i], o_1(x_i),o_2(x_i))=((\triangleleft
(z),\triangleright (z)], o_1(z),o_2(z))$.

Now, we take a representative $z_1$ of the remaining atomic class
of $C$ such that $Z=\{z_1,z_2,\ldots,z_k\}$ a complete set of
representatives of the $k$ distinct atomic classes. We rearrange
the order of elements in $Z$ such that $Z=\{z_1,z_2,\ldots,z_k\}$
is a complete set of atomic class representatives in row echelon
form. By Theorem 3.3, $Z=\{z_1,z_2,\ldots,z_k\}$ is a biproper
$p$-basis for $C$ in row echelon form.

 Suppose that
$0\not\in[x_1]$. Then $[x_1]=(\triangleleft (x_1),\triangleright
(x_1)]$. Since $Z$ is a biproper $p$-basis for $C$, the codeword
$x_1$ can be expressed as a $p$-linear combination of the elements
of $Z$, that is, $$x_1=\sum_{1\leq j\leq k}a_{j}z_j \eqno(12)$$
where $a_{j}\in Z_p$. Since $Z=\{z_1,z_2,\ldots,z_k\}$ is a
biproper $p$-basis for $C$ in row echelon form, it follows that if
$a_{j_1}$ is the first nonzero coefficient in the expression of
$x_1$, then $\triangleleft (x_1)=\triangleleft (z_{j_1})$,
$o_1(x_1)=o_1(z_{j_1})$ and $\triangleright
(x_1)\geq\triangleright (z_{j_1})$.

 (1) $\triangleright
(x_1)=\triangleright (z_{j_1})$. In this case, if
$o_2(x_i)=o_2(y_{j_1})$, we are done. Otherwise,
$o_2(x_1)>o_2(z_{j_1})$. Then there exists $j_2$, $a_{j_2}\neq 0$
in (12), $j_1<j_2\leq k$, such that $\triangleright
(x_1)=\triangleright (z_{j_2})$, $o_2(x_1)=o_2(z_{j_2})$ and
 $\triangleleft (x_1)\leq\triangleleft (z_{j_2})$. Suppose that $\triangleleft
(x_1)=\triangleleft (z_{j_2})$ and $o_1(x_1)>o_1(z_{j_2})$. Since
that $\triangleleft (x_1)=\triangleleft (z_{j_1})$ and
$o_1(x_1)=o_1(z_{j_1})$, it follows that there exists a unit $a$
in $Z_{p^\alpha}$ such that $s=az_{j_1}-x_1$
 ends earlier than $x_1$. Then $s$ can be
expressed as the $p$-linear combination that uses codewords of
 $Y=\{y_1,y_2,\ldots,y_k\}$ later than $z_{j_2}$ and $az_{j_1}=x_1+s$ can be
expressed as the $p$-linear combination that uses $z_{j_1}$,
$z_{j_2}$ and codewords of
 $Y=\{y_1,y_2,\ldots,y_k\}$ later than $z_{j_2}$. Then $\triangleleft
(az_{j_1})=\triangleleft (z_{j_2})$ and $o_2(az_{j_1})\geq
o_2(z_{j_2})$, which is a contradiction. Therefore, $\triangleleft
(x_1)<\triangleleft (z_{j_2})$. Since $X=\{x_1,x_2,\ldots,x_k\}$
is a $p$-basis for a linear code $C$ over $Z_{p^\alpha}$, it
follows that $z_{j_2}\in C$ can be expressed as the $p$-linear
combination
 $$z_{j_2}=c_1x_1+c_2x_2+\cdots+c_kx_k$$
  for some
$c_1,c_2,\ldots,c_k\in Z_p$. If $c_1\neq 0$, then $\{z_{j_2},
x_2,\ldots,x_k\}$ is a $p$-basis for $C$, and, therefore, the
trellis $T'=T_{z_{j_2}}\times T_{x_2}\times \cdots \times T_{x_k}$
represents $C$. But since $\triangleright (x_1)=\triangleright
(z_{j_2})$, $o_2(x_1)=o_2(z_{j_2})$ and $\triangleleft
(x_1)<\triangleleft (z_{j_2})$, it follows that $[z_{j_2}]\subset
[x_1]$,
 and, therefore, $T_{z_{j_2}}\prec_{\Theta}T_{x_{1}}$. Therefore,
$T'\prec_{\Theta}T$, and $T$ is not
 $\prec_{\Theta}$ minimal. Hence, $c_1=0$. Since $\triangleright (x_1)=\triangleright
(z_{j_2})$ and $o_2(x_1)=o_2(z_{j_2})$ , it follows that there
exists a unit $c$ in $
 Z_{p^\alpha}$ such that $t=z_{j_2}+cx_1
 =c_{2}x_{2}+\cdots+c_kx_k+cx_1$ ends earlier than $z_{j_2}$.
 By Lemma 6.1 [20], $t$ can be
expressed as the $p$-linear combination of $x_1$ and later
codewords of $X=\{x_1,x_2,\ldots,x_k\}$. Hence we have a
 $p$-basis $\{t,x_2,\ldots,x_k\}$
 for $C$, and, therefore, the
trellis $T''=T_{t}\times T_{x_{2}}
   \times\cdots\times T_{x_k}$
represents $C$. But since $\triangleleft(t)=\triangleleft (x_1)$,
and $\triangleright (t)<\triangleright (x_1)$, it follows that
$[t]\subset [x_{1}]$ and $T_t\prec_{\Theta}T_{x_{1}}$. Therefore,
$T''\prec_{\Theta}T$, and $T$ is not $\prec_{\Theta}$ minimal.
Thus, it leads to
 a contradiction.

(2) $\triangleright (x_1)>\triangleright (z_{j_1})$. Since
$X=\{x_1,x_2,\ldots,x_k\}$ is a $p$-basis for a linear code $C$
over $Z_{p^\alpha}$, it follows that $z_{j_1}\in C$ can be
expressed as the $p$-linear combination
$$z_{j_1}=b_1x_1+b_2x_2+\cdots+b_kx_k$$
 for some
$b_1,b_2,\ldots,b_k\in Z_p$. If $b_1\neq 0$, then $\{z_{j_1},
x_2,\ldots,x_k\}$ is a $p$-basis for $C$, and, therefore, the
trellis $T'=T_{z_{j_1}}\times T_{x_2}
   \times \cdots\times T_{x_k}$
represents $C$. But since $\triangleleft (x_1)=\triangleleft
(z_{j_1})$, and $\triangleright (x_1)>\triangleright (z_{j_1})$,
it follows that $[z_{j_1}]\subset [x_{1}]$ and
$T_{z_{j_1}}\prec_{\Theta}T_{x_{1}}$. Therefore,
$T'\prec_{\Theta}T$, and $T$ is not
 $\prec_{\Theta}$ minimal. Hence, $b_1=0$. Since $\triangleleft (x_1)=\triangleleft
 (z_{j_1})$ and $o_1(x_1)=o_1(z_{j_1})$,  there exists a unit $b$ in $
 Z_{p^\alpha}$ such that $u=z_{j_1}+bx_1=b_2x_2+\cdots+b_kx_k+bx_1$
  starts later than $z_{j_1}$.  By Lemma 6.1 [20], $u$ can be
expressed as the $p$-linear combination of $x_1$ and later
codewords of $X=\{x_1,x_2,\ldots,x_k\}$. Hence we have a
 $p$-basis $\{u,x_2,\ldots,x_k\}$
 for $C$, and, therefore, the
trellis $T''=T_u\times T_{x_2}\times \cdots \times T_{x_k}$
represents $C$. But since $\triangleleft(u)\geq\triangleleft
(z_{j_1})+1=\triangleleft (x_1)+1$, and $\triangleright
(u)=\triangleright (x_1)$, it follows that $[u]\subset [x_{1}]$
and $T_u\prec_{\Theta}T_{x_{1}}$. Therefore, $T''\prec_{\Theta}T$,
and $T$ is not $\prec_{\Theta}$ minimal. Thus, it leads to
 a contradiction.

Therefore,
$([x_1],o_1(x_1),o_2(x_1))=((\triangleleft(z_{j_1}),\triangleright
  (z_{j_1})],\\o_1(z_{j_1}),o_2(z_{j_1}))$. Hence, for each $1\leq i\leq k$,
  either $0\in[x_i]$ or there
exists $z\in Z$ such that $x_i$ belongs to the same equivalence
class of $C$. Since $z$ is atomic, it follows that there exists
$y\in Y$ such that $z$ belongs to the same equivalence class of $C$.
Thus, for each $1\leq i\leq k$, either $0\in[x_i]$ or there exists
$y\in Y$ such that $x_i$ belongs to the same equivalence class of
$C$.
 \end{proof}

\medskip
 Next, we generalize the result of Theorem 4.1. We let
$\sigma_j(\cdot)$ denote a cyclic shift to the left $j$ times, and
consider the corresponding cyclic shift of $C$, namely,
$C_j=\sigma_j(C)$. Let $\rho_j(\cdot)$ denote a cyclic shift to
the right $j$ times. \linebreak
 For $x=(x_0,x_1,\ldots,x_k)\in
Z_{p^\alpha}$, $\rho_j(x)=(x_{n-j},\ldots, x_{n-1},\linebreak
x_0,x_1,\ldots, x_{n-1-j})$. Let $[x]=(a,b]$, a span of $x$,
define the action of $\rho_j$ on a quadruple as follows:
\begin{eqnarray}
\rho_j((x,[x],o(x_a),o(x_b)))=
(\rho_j(x),(a+j,b+j],& &\nonumber\\
o(\rho_j(x)_{a+j}),
 o(\rho_j(x)_{b+j})),& &
 \end{eqnarray}
 where $o(\rho_j(x)_{a+j})$ is the order of the $(a+j)$-th
component of $\rho_j(x)$ in $Z_{p^\alpha}$. Clearly,
$x_a=\rho_j(x)_{a+j}$,
$x_b=\rho_j(x)_{b+j}$ and $\rho_j((x,[x],o(x_a),o(x_b)))=(\rho_j(x),(a+j,b+j],o(x_a),o(x_b))$.\\

\begin{theorem}
  Let $X=\{x_1,x_2,\ldots,x_k\}$ be a $p$-basis for a linear
code $C$ of length $n$ and $p$-dimension $k$ over $Z_{p^\alpha}$.
Let $j$ be any integer in the set
${\mathcal{I}}=\{0,1,\ldots,n-1\}$. Let
$Y_j=\{y_1,y_2,\ldots,y_k\}$ be a biproper $p$-basis for
$C_j=\sigma_j(C)$ in row echelon form. Let $T=T_{x_1}
   \times\cdots\times T_{x_k}$ is a $\prec_{\Theta}$ minimal
linear tail-biting trellis for $C$. Then for each $1\leq i\leq k$,
either $j\in[x_i]$ or there exists $y\in Y_j$ such that
$\sigma_j(x_i)$ belong to the same equivalence class of $C_j$,
namely, $([x_i], o(x_a),o(x_b))=((\triangleleft
(y)+j,\triangleright (y)+j], o_1(y),o_2(y))$, where $[x_i]=(a,b]$.
\end{theorem}
\begin{proof}
 This follows immediately by combining
Theorem 4.1 and Proposition 3.2 in [12].
 \end{proof}

\medskip
 Now, we can generalize the definition of characteristic
generators for a linear code over a field to  a linear code $C$
over $Z_{p^\alpha}$. Theorem 4.2 makes it possible to characterize
all the $\prec_{\Theta}$ minimal linear tail-biting trellises for
a linear code $C$ over $Z_{p^\alpha}$ in terms of a small set of
characteristic generators. For each $j\in \mathcal{I}$, let $Y_j$
denote the lexicographically first biproper $p$-basis for
$C_j=\sigma_j(C)$
 in row echelon form, define $X_j=\rho_j(Y_j)$. Then $X_j$ is a subset of $C$ for all
 $j$.\\

\begin{definition}
 Let $C$ be a linear code of
length $n$ over $Z_{p^\alpha}$. A \emph{characteristic generator
}for $C$ is a quadruple consisting of a codeword
$x=(x_0,x_1,\ldots,x_{n-1})\in C$, a interval $[x]=(a,b]$ such
that $x_a,x_b$ are nonzero, $ o(x_a)$ and $o(x_b)$. We also let
$Y_j=\{(y,(c,d],o(y_c),o(y_d))\mid y\in Y_j, (c,d]$ is the
conventional span of $y$\}. The set of all the characteristic
generators for $C$ is given by
\begin{eqnarray}
X=X_0\cup X_1\cup\cdots \cup X_{n-1}\qquad\qquad& & \nonumber \\
=Y_0\cup \rho_1(Y_1)\cup\cdots \cup \rho_{n-1}(Y_{n-1})\,& &
\end{eqnarray}
 with the understanding that $[x]=(\triangleleft (y)+j,\triangleright
(y)+j]$, $o(x_a)=o_1(y)$ and $o(x_b)=o_2(y)$ for each $x\in X_j$,
where $y=\sigma_j(x), [x]=(a,b]$. The \emph{characteristic matrix}
$\mathcal{X}$ for $C$ is the $|X|\times n$ matrix having the
elements of $X$ as
its rows.\\
\end{definition}

 \begin{theorem}
  Let $X=\{x_1,x_2,\ldots,x_k\}$ be a $p$-basis for a linear
code $C$ of length $n$ and $p$-dimension $k$ over $Z_{p^\alpha}$.
If $T=T_{x_1}\times T_{x_2}
   \times\cdots\times T_{x_k}$ is a $\prec_{\Theta}$ minimal linear
    tail-biting trellis for $C$, then the trellis $T$ can be constructed as
$$T=T_{z_1}\times T_{z_2}
   \times\cdots\times T_{z_k}$$ where $z_1,z_2,\ldots,z_k$ are $k$ $p$-linearly
independent characteristic generators for
   $C$.
   \end{theorem}
\begin{proof}
 It is obvious that $[x_i]\neq
 \mathcal{I}$ for all $i=1,2,\ldots,k$. Otherwise, we can replace
 $T_{x_i}$ by a strictly smaller trellis $T'_{x_i}$ for $x_i$
 using the span $[x_i]=(0,n-1]$. Hence, there exists at least one
 $j\in \mathcal{I}$ such that $j \notin [x_i]$.

We proceed by induction on $p$-dimension $k$ of $C$. If $k=1$,
there exists $y\in Y_j$
 such that $\sigma_j(x_1)$ belong to the same equivalence class of $C_j$, namely, $([x_1],
o(x_a),o(x_b))=((\triangleleft (y)+j,\triangleright (y)+j],
o_1(y),o_2(y))$, where $[x_1]=(a,b]$. In the case,
  $\rho_j(y)$ is a $p$-basis, and, therefore, the trellis
$T'= T_{\rho_j(y)}$ represents $C$. Then  $T=T'=T_{\rho_j(y)}$.
The result is proved. Suppose that $k>1$. It is easy to show that
$\{x_2,\ldots,x_k\}$ is a $p$-linearly independent $p$-generator
sequence. Let $C'=p$-span($\{x_2,\ldots,x_k\}$) be a subcode of
$C$. It is obvious that $T'=T_{x_2}
   \times\cdots\times T_{x_k}$ is a $\prec_{\Theta}$ minimal
linear tail-biting trellis for $C'$. By definition the
$p$-dimension of $C'$ is $k-1$. Then by induction the trellis $T'$
can be constructed as
$$T'= T_{z_2}\times\cdots\times T_{z_k}$$
    where $z_2,\ldots,z_k$ are $k-1$ $p$-linearly
independent characteristic generators for
   $C'$. It is easy to show that $z_2,\ldots,z_k$ are also $k-1$ $p$-linearly
independent characteristic generators and  $x_1,z_2,\ldots,z_k$ is
also a $p$-basis for $C$.

 Now, there exists at least one
 $j\in \mathcal{I}$ such that $j \notin [x_1]$. This implies that
 there exists $y\in Y_j$ such that $\sigma_j(x_1)$
belong to the same equivalence class of $C_j$, namely, $([x_1],
o(x_a),o(x_b))=((\triangleleft (y)+j,\triangleright (y)+j],
o_1(y),o_2(y))$, where $[x_1]=(a,b]$. Since $x_1,z_2,\ldots,z_k$
is a $p$-basis for $C$, it follows that $\rho_j(y)$ can be
expressed as a $p$-linear combination
$$\rho_j(y)=a_1x_1+a_2z_2+\cdots+a_kz_k,$$
 where $a_{j}\in Z_p$. If
$a_1=0$, then there exists a unit $u$ in $
 Z_{p^\alpha}$ such that the span of $t=\rho_j(y)+ux_1
 =a_{2}z_{2}+\cdots+a_kz_k+ux_1$ starts later than $x_1$. By Lemma 6.1 [20], $t$ can be
expressed as the $p$-linear combination of $x_1$ and later
codewords of $\{x_1,z_2,\ldots,z_k\}$. Hence we have a
 $p$-basis $\{t,z_2,\ldots,z_k\}$
 for $C$, and, therefore, the
trellis $T'=T_{t}\times T_{z_{2}}
   \times\cdots\times T_{z_k}$
represents $C$. But since $([x_1], o(x_a),o(x_b))=((\triangleleft
(y)+j,\triangleright (y)+j], o_1(y),o_2(y))$, where $[x_1]=(a,b]$,
it follows that $[t]\subset[x_{1}]$ and
$T_t\prec_{\Theta}T_{x_{1}}$. Therefore, $T'\prec_{\Theta}T$, and
$T$ is not $\prec_{\Theta}$ minimal. Thus, it leads to
 a contradiction. Hence,  $a_1\neq0$ and we have a
 $p$-basis $\{\rho_j(y),z_2,\ldots,z_k\}$
 for $C$, and, therefore, the trellis
$T''=T_{\rho_j(y)}\times T_{z_2}
   \times\cdots\times T_{z_k}$ represents $C$. Then
   $T=T''=T_{\rho_j(y)}\times T_{z_2}
   \times\cdots\times T_{z_k}$, where $\{\rho_j(y),z_2,\ldots,z_k\}$ are $k$ $p$-linearly
independent characteristic generators for
   $C$.
   \end{proof}

\medskip
We review a key theorem in [13] as follows:\\
\begin{lemma}
Let $T$ be a group tail-biting trellis over an abelian group. Then
$T$ can be factored as $T=T_{x_1}\times T_{x_2}
   \times\cdots\times T_{x_k}$, where $T_{x_i}$
   is elementary trellis over a group, for $i=1,2,\ldots,k$.
 \end{lemma}

\medskip
Now, we have the following.\\
\begin{theorem}
Let $T$ be a linear tail-biting trellis over $Z_{p^\alpha}$. Then
$T$ can be factored as $T=T_{x_1}\times T_{px_1}\times\cdots\times
T_{p^{\alpha_1}x_1}\times T_{x_2}\times T_{px_2}\times\cdots\times
T_{p^{\alpha_2}x_2}
   \times\cdots\times T_{x_k}\times T_{px_k}\times\cdots\times
T_{p^{\alpha_k}x_k}$, where $T_{x_i},T_{px_i},\ldots,
T_{p^{\alpha_i}x_i}$ are elementary trellises over $Z_{p^\alpha}$,
   and the order of $x_i$ is $p^{\alpha_i}$, for $i=1,2,\ldots,k$.
\end{theorem}

\begin{proof}
Obviously, $T$ is also a group tail-biting trellis over the
additive group of $Z_{p^\alpha}$. By Lemma 4.1, as a group
tail-biting trellis, $T$ can be factored as $T=T_{x_1}\times
T_{x_2}\times\cdots\times T_{x_k}$, where $T_{x_1},T_{x_2},
   \ldots,T_{x_k}$ are elementary trellises over a group. For
   $i=1,2,\ldots,k$, the elementary tail-biting trellis $T_{x_i}$
   is a minimal group tail-biting trellis for group code $\langle x_i\rangle$  over the
additive group of $Z_{p^\alpha}$, then $T_{x_i}$ is also a minimal
linear tail-biting trellis for linear code $\langle x_i\rangle$
over $Z_{p^\alpha}$. Let the order of $x_i$ is $p^{\alpha_i}$.
Since the linear code $\langle x_i\rangle$ have a $p$-basis
$x_i,px_i,\ldots,p^{\alpha_i}x_i$, $T_{x_i}=T_{x_i}\times
T_{px_i}\times\cdots\times T_{p^{\alpha_i}x_i}$.
 \end{proof}

\medskip
 \begin{theorem}
Every linear tail-biting trellis $T$ for a linear code $C$ of
length $n$ and $p$-dimension $k$ over $Z_{p^\alpha}$ that is
minimal under $\prec_{\Theta}$ can be constructed as
$$T=T_{z_1}\times T_{z_2}
   \times\cdots\times T_{z_k}$$
where $z_1,z_2,\ldots,z_k$ are $k$ $p$-linearly independent
characteristic generators for
   $C$.
 \end{theorem}

\begin{proof}
By Theorem 4.4, the trellis $T$ can be factored as
$T=T_{x_1}\times T_{px_1}\times\cdots\times
T_{p^{\alpha_1}x_1}\times T_{x_2}\times T_{px_2}\times\cdots\times
T_{p^{\alpha_2}x_2}
   \times\cdots\times T_{x_m}\times T_{px_m}\times\cdots\times
T_{p^{\alpha_m}x_m}$, where $T_{x_i},T_{px_i},\ldots,
T_{p^{\alpha_i}x_i}$ are elementary trellises over $Z_{p^\alpha}$,
   and the order of $x_i$ is $p^{\alpha_i}$, for $i=1,2,\ldots,m$.
   If $\alpha_1+\alpha_2+\cdots+\alpha_m >k$, then $x_1,\;px_1,\ldots
   p^{\alpha_1}x_1,\;x_2,\;px_2,\ldots,p^{\alpha_2}x_2,\ldots,\;x_m,\;px_m,\ldots,\linebreak
   p^{\alpha_m}x_m$
   are $p$-linearly dependent. Suppose that $p^{j}x_i$ can be \linebreak
   expressed as a $p$-linear combination of
   $p^{j+1}x_i,\ldots,x_m,px_m,\linebreak \ldots,p^{\alpha_m}x_m$, where
   $j\neq \alpha_m$ or $i\neq m$. Therefore,
   $T'=T_{x_1}\times T_{px_1}\times\cdots\times
T_{p^{\alpha_1}x_1}\times T_{x_2}\times T_{px_2}\times\cdots\times
T_{p^{\alpha_2}x_2}
   \times\cdots \times T_{p^{j-1}x_i} \times T_{p^{j+1}x_i}\times\cdots\times T_{x_m}\times T_{px_m}\times\cdots\times
T_{p^{\alpha_m}x_m}$ is a linear trellis such that $C(T)=C(T')$
and $T'\prec_{\Theta}T$.\linebreak
 Hence $\alpha_1+\alpha_2+\cdots+\alpha_m \leq k$ and $x_1,\;px_1,\ldots
   p^{\alpha_1}x_1,\;x_2,\linebreak
   px_2,\ldots,p^{\alpha_2}x_2,\ldots,\;x_m,\;px_m,\ldots,
   p^{\alpha_m}x_m$
   are $p$-linearly independent. Clearly, $C(T)=C$ if and only if $C=p$-span($\{x_1,\;px_1,\ldots
   p^{\alpha_1}x_1,\;x_2,\;px_2,\ldots,p^{\alpha_2}x_2,\ldots,\;x_m,\;px_m,\linebreak\ldots,
   p^{\alpha_m}x_m\}$), which implies $\alpha_1+\alpha_2+\cdots+\alpha_m=k$ and $\{x_1,\;px_1,\ldots
   p^{\alpha_1}x_1,\;x_2,
   px_2,\ldots,p^{\alpha_2}x_2,\ldots,\;x_m,\;px_m,\ldots,\linebreak
   p^{\alpha_m}x_m\}$ is a $p$-basis for $C$. By Theorem 4.3, the trellis $T$ can be constructed as
$$T=T_{z_1}\times T_{z_2}
   \times\cdots\times T_{z_k}$$ where $z_1,z_2,\ldots,z_k$ are $k$ $p$-linearly
independent characteristic generators for
   $C$.
\end{proof}

\medskip
A similar argument makes it possible to characterize minimal
linear tail-biting
  trellises under any of the minimality orders defined in (4)-(9) for a linear code
  $C$ over $Z_{p^\alpha}$.\\

  \begin{theorem}
   Every linear tail-biting trellis $T$ for a linear code $C$ of
length $n$ and $p$-dimension $k$ over $Z_{p^\alpha}$ that is
minimal under any of the minimality orders defined in (4)-(9) can
be constructed as
$$T=T_{z_1}\times T_{z_2}
   \times\cdots\times T_{z_k}$$
where $z_1,z_2,\ldots,z_k$ are $k$ $p$-linearly independent
characteristic generators for
   $C$.
 \end{theorem}
\begin{proof}
 It is obvious that the set of minimal
 trellises with respect to these three orders,  the vertex-product order
 $\prec_{\Pi}$, the vertex-max order $\prec_{max}$, and the vertex-sum order
 $\prec_{\Sigma}$, is a subset of the set of $\prec_{\Theta}$-minimal trellises, and the result follows
 directly from Theorem 4.5. Similarly, with respect to the other three orders $\prec_{\Pi\mathcal{E}}$, $\prec_{max\mathcal{E}}$
 and $\prec_{\Sigma\mathcal{E}}$ introduced in (7)-(9), the set of minimal
 trellises is a subset of the set of minimal trellises under the order $\prec_{\mathcal{E}}$ defined by the edge-class profile
 $(|E_0|,|E_1|,\ldots,|E_{n-1}|)$ of the trellis. Observe that the
 proof of Theorem 4.1(and, hence, also of Theorem 4.2, Theorem 4.3 and Theorem 4.5)
 holds without change if we replace ``$\prec_{\Theta}$" with
 ``$\prec_{\mathcal{E}}$" throughout. Thus, Theorem 4.5 holds for $\prec_{\mathcal{E}}$-minimal
 trellises as well, and the claim with respect to the $\prec_{\Pi\mathcal{E}}$, $\prec_{max\mathcal{E}}$
 and $\prec_{\Sigma\mathcal{E}}$ orders follows.
 \end{proof}

\medskip
Now we go back to the codes over finite abelian groups.
Let $G$ be a finite abelian group and $C$ be a group code of
length $n$ over $G$. Then $G$ can decomposed into a direct product
of cyclic groups, i.~e.,
 $$ G=( Q_{p_1^{\alpha_{11}}}+\cdots+Q_{p_1^{\alpha_{1m_1}}})+\cdots+ ( Q_{p_r^{\alpha_{r1}}}+
 \cdots+Q_{p_r^{\alpha_{rm_r}}}),$$
where $p_1,\ldots, p_r$ are distinct primes. We can decompose $C$
into $r$ codes $C_1,\ldots,C_r$, i.e., $C=C_1+\cdots+C_r$.
  Hence, it is sufficient to consider the case of a code over a $p$-group $H$, i.e.,
  $H= Q_{p^{\alpha_1}}+\cdots+Q_{p^{\alpha_m}}$, where $\alpha_1\leq \cdots\leq \alpha_m$.
  Then by Lemma 8.5 in [20] a code
of length $n$ over $p$-group $H$ is equivalent to a linear code
   of length $mn$ over $Z_{p^{\alpha_m}}$.
  If we denote the minimal trellis for a code $C$ over a finite abelian group $G$ by $T$,
   the minimal trellis for the code $C_i$ over $p_i$-group $H_i$ by $T_i$, then $T=T_1
   \times\cdots\times T_r$. It is easy to see that $T_i$ is a trellis that
   can be obtained by sectionalizing a minimal trellis for linear code
    $S_i$ of length $m_in$ over $Z_{p_i^{\alpha_{im_i}}}$, where the
    code $S_i$ is equivalent to the code $C_i$ over $p_i$-group $H_i$.

We can obtain a generator matrix $A_i$ for $C_i$ over $p_i$-group
$H_i$ from a generator matrix $A$ for $C$, where the element in
matrix $A_i$ in the $j$-th row and the $l$-th
 column is $a_{jl}^i$ if $a_{jl}$ in matrix $A$ in the $j$-th row and the $l$-th
 column can decomposed into a direct product of
 $a_{jl}^1, a_{jl}^2, \ldots, a_{jl}^r$, $a_{jl}^i\in H_i$ for
 $i=1,2,\ldots,r$. We also can obtain a generator matrix $B_i$ for
 the linear code $S_i$ of length $m_in$ over
 $Z_{p_i^{\alpha_{im_i}}}$ from $A_i$ by Lemma 8.6 in [20]. By Theorem 4.5, we can
 construct a minimal tail-biting trellis $T'_i$ for $S_i$, for
 $i=1,2,\ldots,r$. Then sectionalizing this trellis will give
 a minimal tail-biting trellis $T_i$ for the code $C_i$ over $p_i$-group
 $H_i$. Thus, the product trellis $T=T_1
   \times\cdots\times T_r$ is a minimal tail-biting trellis for $C$
   over a finite abelian group $G$. Therefore, we show that
although minimal tail-biting trellises for group codes are
generally not unique, every minimal linear tail-biting trellis for
a group code over a finite abelian group necessarily can be
construct from its characteristic matrix.

\section{Computation of the Characteristic Matrix }

Let $C$ be a linear code of length $n$ and $p$-dimension $k$ over
$Z_{p^\alpha}$. The basic properties of characteristic generators
for $C$ are investigated in [30]. It was proved that the spans of
any two characteristic generators for $C$ do not start at the same
position and have the same order of their starting components
simultaneously and end at the same position and have the same
order of their ending components simultaneously as well. The
authors also proved that the number of characteristic generators
for $C$ is $\sum\limits_{i\in \chi(C)}k_i$ and less than $nk$,
where $p^{k_i}$, $0\leq k_i\leq\alpha$, is the maximal order of
the $i$-th components of codewords in $C$.

It follows from [30] that the characteristic matrix $\mathcal{X}$
for $C$ is a $\sum\limits_{i\in \chi(C)}k_i\times n$ matrix. For
simplicity, we henceforth assume that $|\chi(C)|=n$ (otherwise, we
puncture out the all-zero columns of $C$). Then the characteristic
matrix $\mathcal{X}$ is a $\sum\limits_{i=0}^{n-1}k_i\times n$
matrix, and $0<k_i\leq \alpha$ for all $i=0,1,\ldots,n-1$.

We now show how the characteristic matrix $\mathcal{X}$ can be
computed from an arbitrary given $p$-generator sequence
$V=\{v_1,v_2,\cdots, v_k\}$ for the module $C$ over
$Z_{p^\alpha}$. The first step is to convert this $p$-generator
sequence into a biproper $p$-basis in row echelon form. As is well
known [20], this can be easily
accomplished by Gaussian elimination, as follows:\\
\begin{quote}
\textsf{Algorithm A}

\quad  \textsf{1) $S\longleftarrow V$.}

\quad  \textsf{2) While there is a nonzero element in $S$ do:}

\quad  \textsf{3) Find $S'\subseteq S$, the set of elements of $S$
having the earliest starting position.}

\quad  \textsf{4) Find $S''\subseteq S'$, the set of elements of
$S'$ having the highest order starting component. }

\quad \textsf{5) pick the last element $v\in S''$, output it, and
set $S\longleftarrow S-\{v\}$.}

\quad\textsf{6) For each remaining $u\in S''$, replace $u$ in $S$
by $u+av$, where $a\in Z_{p^\alpha}$ is such that $u+av$ starts
laster that $u$.}

\quad \textsf{7) end.}\\
\end{quote}

The algorithm A given above starts with an arbitrary $p$-generator
sequence $V=\{v_1,v_2,\cdots, v_k\}$ for the module $C$ over
$Z_{p^\alpha}$ and finds a proper $p$-generator sequence in row
echelon form, by Lemma 6.10 in [20] the proper $p$-generator
sequence in row echelon form is a $p$-basis. Its running time is
bounded by $\emph{O}(n^3)$ operations over $Z_{p^\alpha}$.\\
\begin{quote}
\textsf{Algorithm B}

\quad \textsf{1)  $S\longleftarrow W$.}

\quad\textsf{2) While  $S$ is not biproper do:}

\quad \textsf{3) Find $S'\subseteq S$, with $|S'|>1$, elements
having the latest ending position, and moreover such that their
ending components are associate;}

\quad \textsf{4) Let $w$ be the last element in $S'$;}

\quad \textsf{5)For each remaining $u\in S'$, replace $u$ in $S$
by $u+aw$, where $a\in Z_{p^\alpha}$ is such that $u+aw$ ends
earlier that $u$.}

\quad \textsf{6) end.}\\
\end{quote}

The algorithm B given above starts with a proper $p$-generator
sequence $W=\{w_1,w_2,\cdots, w_k\}$ in row echelon form for the
module $C$ over $Z_{p^\alpha}$ and finds a biproper $p$-generator
sequence in row echelon form. Its running time is bounded by
$\emph{O}(n^3)$ operations over $Z_{p^\alpha}$. To ensure that the
resulting biproper $p$-basis is the lexicographically first
biproper $p$-basis for $C$ in row echelon form, we need to select
the lexicographically first codeword in each atomic equivalence
class, this can be easily done as follows:\\
\begin{quote}
\textsf{Algorithm C}

 \quad\textsf{for all $i\in\{1,2,\ldots,k\}$ do}

\quad\textsf{while ($\exists x_j$ and $a\in Z_{p^\alpha}$ such
that $[x_j]\subset [x_i]$ and $x_i+ax_j\prec_L x_i$)}

\quad\textsf{do $x_i:=x_i+ax_j$.}\\

\end{quote}

The complexity of Algorithm C is also at most $\emph{O}(n^3)$.
This immediately yields an $\emph{O}(n^4)$ algorithm for the
computation of the characteristic matrix. For each of the $n$
cyclic shifts of $C$, compute the lexicographically first biproper
$p$-basis in row echelon form using Algorithm A, Algorithm B and
Algorithm C, then rotate cyclically to the right and form the set
of characteristic generators $X$ as in (14).

However, we can simplify the computation of the characteristic
matrix. since $C$ has $p^k$ elements and $C$ is a subgroup of
$Z_{p^\alpha}^n$ under the componentwise addition operation of
$Z_{p^\alpha}$, $k_i\leq k$ for all $i=0,1,\ldots,n-1$, and
therefore, for each $j\in \mathcal{I}$, the sets $\rho_j(Y_j)$ and
$\rho_{j+1}(Y_{j+1})$ in (14) have at least $k-k_j$ characteristic
generators in common by Corollary 9 in [30]. Then we obtain an
efficient algorithm for computing the characteristic matrix as
follows:\\

\textbf{Algorithm I}

\textsf{Input:} An arbitrary given $p$-generator sequence for a
linear code $C$ length $n=|\chi(C)|$ and $p$-dimension $k$ over
$Z_{p^\alpha}$.

\textsf{Output:} The set $X$ of $n$ characteristic generators for
$C$.

  1) Using Algorithm A, Algorithm B and Algorithm C,\linebreak
  compute $X_0=\{x_1,x_2,\ldots,x_k\}$, the lexicographically first \linebreak
  biproper $p$-basis in row echelon form. Set $j:=0$. Also set \linebreak
  $X=\{(x_1,[x_1],o_1(x_1),o_2(x_1)),(x_2,[x_2],o_1(x_2),o_2(x_2)),\ldots,\\
  (x_k,[x_k],o_1(x_k),o_2(x_k))\}$,
where $[x_i]=((\triangleleft (x_i),\triangleright (x_i)]$ for all
$i$.

  2) Find the set $T\subseteq \{1,2,\ldots,k\}$ and $|T|=k_j$, such that $\triangleleft
  (x_t)=0$ for each $t\in T$. Then do $x_i=\sigma_1(x_i)$ for all
  $i=1,2,\ldots,k$.

  3) For all $t\in T$ do, while $\exists i\in  \{1,2,\ldots,k\}\setminus\{t\}$
  and $o_2(x_i)<o_2(x_t)$ if $i\in T$ such that
  $\triangleleft(x_i)=\triangleleft(x_t)$ and $o_1(x_i)\linebreak
  =o_1(x_t)$, do $x_t:=x_t+ax_i$,
  where $a\in Z_{p^\alpha}$ is such that $x_t+ax_i$ starts
laster that $x_t$(and go back).

  4) For all $t\in T$ do, while $\exists i\in  \{1,2,\ldots,k\}\setminus \{t\}$,
  $o_2(x_i)<o_2(x_t)$ if $i\in T$
and $a\in Z_{p^\alpha}$ such that
  $(\triangleleft(x_i),\triangleright (x_i)]$ is a proper subset of
  $(\triangleleft(x_t),\triangleright(x_t)]$
  and $x_t+ax_i\prec_L x_t$, do $x_t:=x_t+ax_i$(and go
  back).

  5) Permuting all the elements in $\{x_1,x_2,\ldots,x_k\}$ such that
  they are in row echelon form.

  6) Set $j:=j+1$. If
  $(\rho_j(x_t),(\triangleleft(x_t)+j,\triangleright(x_t)+j],o_1(x_t),\linebreak
  o_2(x_t))$ is not in $x$, add this pair to $X$. If
  $|X|=\sum\limits_{i=0}^{n-1}k_i$, return $X$ and exit; else go
  to the step 2).\\

The complexity of Algorithm I is at most $\emph{O}(n^3)$.

We now show that above Algorithm I is reasonable. First, we prove
that for $j=0,1,\ldots,n-1$, if the set $\{x_1,x_2,\ldots,x_k\}$
is the lexicographically first biproper $p$-basis in row echelon
form of $C_j=\sigma_j(C)$, then, after the execution of the steps
2)-5), the set $\{x_1,x_2,\ldots,x_k\}$ is the lexicographically
first biproper $p$-basis in row echelon form of
$C_{j+1}=\sigma_{j+1}(C)$. Then, by Theorem 8, Theorem 9 in [30]
and the definition of biproper $p$-basis, the fact that
$|\chi(C)|=n$ guarantees that there are $t_1,t_2,
\ldots,t_{k_j}\in \{1,2,\ldots,k\}$ such that
$\triangleleft(x_{t_{i}})=0, \linebreak o_1(x_{t_{i}})=p^i, i=1,2,
\ldots,k_j$. Set $\{t_1,t_2,\ldots,t_{k_j}\}=T$ and $|T|=k_j$.
After the cyclic shift, the set $\{x_1,x_2,\ldots,x_k\}$ is a
$p$-basis of $C_{j+1}$. For all $i\in \{1,2,\ldots,k\}\setminus
T$, the cyclic shift decreases $\triangleleft(x_i)$ and
$\triangleright (x_i)$ by one. Thus all the elements in
$\{x_1,x_2,\ldots,x_k\}\setminus \{x_t|t\in T\}$ either start at
different positions or start the same position but their starting
components have different orders and either end at different
positions or end the same position but their ending components
have different orders. Moreover, the elements of $\{x_t|t\in T\}$
are all elements in $\{x_1,x_2,\ldots,x_k\}$ with the property
that $\triangleright(x_t)=n-1$ and the orders of their ending
components are\linebreak
 $p^1, p^2, \ldots,p^{k_j}$. This property
obviously remains after the execution of the step 3). And this
step makes sure that
$(\triangleleft(x_i),o_1(x_i))\neq(\triangleleft(x_t),o_1(x_t))$
for all $t\in T$ and $i\in \{1,2,\linebreak
\ldots,k\}\setminus\{t\}$. Then the set $\{x_1,x_2,\ldots,x_k\}$
is a biproper $p$-basis of $C_{j+1}$ after the step 3). Each
element in $\{x_1,x_2,\linebreak \ldots,x_k\}\setminus \{x_t|t\in
T\}$ is lexicographically first in its atomic equivalence class
since this property is preserved by the cyclic shift. After the
step 4), each $x_t$ for $t\in T$ is also lexicographically first
in its atomic equivalence class. Therefore,
$\{x_1,x_2,\ldots,x_k\}$ is the lexicographically first biproper
$p$-basis of $C_{j+1}$. After the step 5),
$\{x_1,x_2,\ldots,x_k\}$ is the lexicographically first biproper
$p$-basis in row echelon form of $C_{j+1}$.

 Next, we only need to
prove that Algorithm I eventually terminates with
$|X|=\sum\limits_{i=0}^{n-1}k_i$. Since the elements of
$\{x_t|t\in T\}$ are all elements in $\{x_1,x_2,\ldots,x_k\}$ with
the property that $\triangleright(x_t)=n-1$ and the orders of
their ending components are $p^1, p^2, \ldots,p^{k_j}$,
$\triangleright(x_t)+j=j-1$ and the orders of ending \linebreak
components of the spans of the elements in $\{\rho_j(x_t)|t\in
T\}$ are $p^1,p^2, \ldots,
 p^{k_j}$, for $j=0,1,\ldots,n-1$. Then after at most $n$
 iterations, the set $X$ contains characteristic generators with
 spans ending at every position in $\{0,1,\ldots,n-1\}$ and the orders of ending
components of the spans are $p^1,p^2, \ldots,
 p^{k_j}$ at ending position $j$, for $j=0,1,\ldots,n-1$.
 Therefore, $|X|=\sum\limits_{i=0}^{n-1}k_i$. \\

 \begin{remark}
We can delete the step Algorithm C in 1), 4) and 5), for the
notations of lexicographic order and proper $p$-basis in row
echelon form were introduced for notational convenience only. In
the step 6), add the pair
$(\rho_j(x_t),(\triangleleft(x_t)+j,\triangleright(x_t)+j],o_1(x_t),
  o_2(x_t))$ to $X$ only if $X$ does not contain any
  characteristic generator with $((\triangleleft(x_t)+j,\triangleright(x_t)+j],o_1(x_t),
  o_2(x_t))$. This simplification eliminates a number of operations at each iteration, but
the complexity of the Algorithm I remains at most $\emph{O}(n^3)$.\\
  \end{remark}

\begin{example}
Consider the code over $Z_8$ generated by:
\[ \left(%
\begin{array}{cccc}
  1 & 2 & 1 & 2 \\
  2 & 0 & 4 & 2 \\
  0 & 0 & 4 & 4 \\
\end{array}%
\right). \]

By Lemma 6.4, a $p$-generator sequence of the code is given below.

\[\left(%
\begin{array}{cccc}
  1 & 2 & 1 & 2  \\
  2 & 4 & 2 & 4  \\
  4 & 0 & 4 & 0 \\
  2 & 0 & 4 & 2 \\
  4 & 0 & 0 & 4 \\
  0 & 0 & 4 & 4 \\
\end{array}%
\right)\]

Using Algorithm A, we get a proper $p$-basis in row echelon form
as follows:

\[\left(%
\begin{array}{cccc}
  1 & 2 & 1 & 2  \\
  2 & 4 & 2 & 4 \\
  4 & 0 & 4 & 0 \\
  0 & 4 & 2 & 6 \\
  0 & 0 & 4 & 4 \\
\end{array}%
\right)\]

Using Algorithm B and Algorithm C, we get the lexicographically
first biproper $p$-basis in row echelon form of $C$. In the code,
$\sum\limits_{i=0}^{3}k_i=10$, then Algorithm I terminates in two
iterations. After the steps 2)-5) in Algorithm I, we can get the
lexicographically first biproper $p$-bases in row echelon form of
$C_1,C_2$. The lexicographically first biproper $p$-bases in row
echelon form of $C,C_1,C_2$ are given by

\[Y_0=\left\{
\begin{array}{cccc}
  1 & 6 & 3 & 0  \\
  2 & 4 & 6 & 0 \\
  4 & 0 & 4 & 0 \\
  0 & 4 & 2 & 6 \\
  0 & 0 & 4 & 4 \\
\end{array}
\right\},
 \]

\[Y_1=\left\{
\begin{array}{cccc}
  2 & 1 & 2 & 1  \\
  4 & 2 & 6 & 0 \\
  0 & 4 & 4 & 0 \\
  0 & 0 & 2 & 6\\
  0 & 0 & 4 & 4 \\
\end{array}
\right\} ,\quad Y_2=\left\{
\begin{array}{cccc}
  1 & 2 & 1 & 2  \\
  2 & 0 & 6 & 4 \\
  4 & 4 & 0 & 0  \\
  0 & 2 & 6 & 0 \\
  0 & 4 & 4 & 0 \\
\end{array}
\right\}
\]
respectively. We get the characteristic matrix for $C$ as follows:

\[ \mathcal{X}=\left(\begin{array}{c}
  x_1   \\
  x_2   \\
  x_3   \\
  x_4   \\
  x_5   \\
  x_6   \\
  x_7   \\
  x_8   \\
  x_9   \\
  x_{10}\\
\end{array}
\right)
 =
 \left(
\begin{array}{cccc}
  1 & 6 & 3 & 0  \\
  2 & 4 & 6 & 0  \\
  4 & 0 & 4 & 0  \\
  0 & 4 & 2 & 6  \\
  0 & 0 & 4 & 4  \\
  1 & 2 & 1 & 2  \\
  6 & 0 & 0 & 2  \\
  4 & 0 & 0 & 4  \\
  1 & 2 & 1 & 2  \\
  6 & 4 & 2 & 0  \\
  \end{array}
\right)
\begin{array}{c}
  ((0,2],3,3) \\
  ((0,2],2,2) \\
  ((0,2],1,1) \\
  ((1,3],1,2) \\
  ((2,3],1,1) \\
  ((1,0],2,3) \\
  ((3,0],2,2) \\
  ((3,0],1,1) \\
  ((2,1],3,2) \\
  ((2,1],2,1) \\
\end{array}\]

Consider the minimal conventional trellises for $C$, $C_1$, $C_2$,
$C_3$, we obtain $4$ minimal tail-biting trellises for $C$ after
an obvious permutation. By the Theorem 4.5, the set $X$ of
characteristic generators is the smallest set of generators from
which all the $\prec_{\Theta}$-minimal tail-biting trellises for
$C$ can be constructed. However, not every choice of $5$
characteristic generators in $X$ produces a
$\prec_{\Theta}$-minimal tail-biting trellis for $C$ under the
product construction. It is necessary for producing a
$\prec_{\Theta}$-minimal tail-biting trellis under the product
construction to satisfy the condition that $5$ characteristic
generators for $C$ are $p$-linearly independent.\\

\end{example}

Now we go back to the codes over finite abelian groups. Since a
code over a finite abelian group can decomposed into a direct
product of codes over those abelian $p$-groups which are all Sylow
$p$-subgroup of the group and a code of length $n$ over $p$-group
is equivalent to a linear code
   of length $mn$ over $Z_{p^{\alpha}}$, by using sectionalization and Algorithm I, we get an efficient algorithm
   for constructing the minimal
tail-biting trellis of a group code over a finite abelian group,
given a generator matrix, see the section IV.

\section{Conclusion}

In this paper, we gives a general solution to the problem of
constructing minimal linear tail-biting trellises for group codes
over finite abelian groups. Thus an important application of our
algorithms is to the construction of minimal trellises for
lattices and some famous nonlinear binary codes, including
Kerdock, Preparata, and Goethals codes.

A research direction worth investigating is using the special
structure of minimal tail-biting trellises for group codes to obtain
faster decoding algorithms.

\section*{Acknowledgment}

The authors would like to thank...

\end{document}